\documentclass[%
aps,
pra,%
superscriptaddress,
amsmath,amssymb,
reprint,%
]{revtex4-1}
\usepackage{graphicx}
\usepackage{dcolumn}
\usepackage{bm}
\usepackage{braket}
\usepackage{color}
\usepackage{hyperref}

\begin{document}

\title{Orbital optimized unitary coupled cluster theory for quantum computer}

\author{Wataru Mizukami}
\email{wataru.mizukami.857@qiqb.otri.osaka-u.ac.jp}
\affiliation{Graduate School of Engineering Science, Osaka University, 1-3 Machikaneyama, Toyonaka, Osaka 560-8531, Japan.}
\affiliation{Center for Quantum Information and Quantum Biology, Institute for Open and Transdisciplinary Research Initiatives, Osaka University, Osaka 560-8531, Japan}
\affiliation{JST, PRESTO, 4-1-8 Honcho, Kawaguchi, Saitama 332-0012, Japan}

\author{Kosuke Mitarai}
\email{mitarai@qc.ee.es.osaka-u.ac.jp}
\affiliation{Graduate School of Engineering Science, Osaka University, 1-3 Machikaneyama, Toyonaka, Osaka 560-8531, Japan.}
\affiliation{QunaSys Inc., Aqua Hakusan Building 9F, 1-13-7 Hakusan, Bunkyo, Tokyo 113-0001, Japan}

\author{Yuya O. Nakagawa}
\affiliation{QunaSys Inc., Aqua Hakusan Building 9F, 1-13-7 Hakusan, Bunkyo, Tokyo 113-0001, Japan}

\author{Takahiro Yamamoto}
\affiliation{QunaSys Inc., Aqua Hakusan Building 9F, 1-13-7 Hakusan, Bunkyo, Tokyo 113-0001, Japan}

\author{Tennin Yan}
\affiliation{QunaSys Inc., Aqua Hakusan Building 9F, 1-13-7 Hakusan, Bunkyo, Tokyo 113-0001, Japan}

\author{Yu-ya Ohnishi}
\email{Yuuya\_Oonishi@jsr.co.jp}
\affiliation{Materials Informatics Initiative, Yokkaichi Research Center, JSR Corporation, 100 Kawajiri-cho, Yokkaichi, Mie, 510-8552, Japan}


\begin{abstract}
We propose an orbital optimized method for unitary coupled cluster theory (OO-UCC) within the variational quantum eigensolver (VQE) framework for quantum computers.
OO-UCC variationally determines the coupled cluster amplitudes and also molecular orbital coefficients.
Owing to its fully variational nature, first-order properties are readily available. 
This feature allows the optimization of molecular structures in VQE without solving any additional equations. 
Furthermore, the method requires smaller active space and shallower quantum circuit than UCC to achieve the same accuracy.
We present numerical examples of OO-UCC using quantum simulators, which include the geometry optimization of the water and ammonia molecules using analytical first derivatives of the VQE.
\end{abstract}

\maketitle

\section{\label{sec:intro} Introduction}
Coupled cluster theory (CC) is one of the most representative electron correlation methods in quantum chemistry~\cite{Bartlett2007RMP,Shavitt2009,Bartlett2012WIRCMS}.
It has some vital features to describe molecular electronic structures reliably.
CC is size-extensive and can be improved systematically by increasing the excitation level.
It converges to full configuration interactions (FCI) faster than truncated configuration interactions (CI) or M{\o}ller-Plesset perturbation theory.
Furthermore, its energy is invariant to unitary transformations among the occupied/virtual orbitals.
It is well known that, for an electronic state where the mean-field approximation works well, the CC models considering up to triple excitations can provide the chemical accuracy (e.g., 1 kcal/mol), if sufficiently large basis functions are employed.
In particular, it has been established that CC singles and doubles with perturbative triples (i.e., CCSD (T)), which incorporates three-body interactions perturbatively, is a highly accurate method.
It is often called the golden standard of molecular electronic structure theory~\cite{Raghavachari1989CPL}.

In the framework of the traditional CC (TCC), the wave-function parameters are determined by solving projected amplitude equations and not variationally.
Owing to its non-variational properties, the validity of TCC is strongly dependent on the reference Hartree--Fock wave function.
Indeed, CCSD and CCSD(T) often suffer from breakdowns in the systems where static electron correlations are strong, for example, multiple-chemical-bond breaking systems~\cite{Bartlett2007RMP}.

This issue can be solved if a CC wave function is parametrized variationally~\cite{Troy2000JCP}.
Some benchmark studies showed that the difference between the variational CC (VCC) and TCC is small for weakly-correlated regions~\cite{Troy2000JCP, Cooper2010JCP, Evangelista2011JCP}.
Nonetheless, VCC has a factorial-scaling in the computational-cost.
Therefore, it is only applicable to a tiny system where FCI can be performed.
However, it has recently been shown that a variant of VCC---unitary coupled cluster (UCC)~\cite{Kutzelnigg1982JCP, Kutzelnigg1983JCP, Kutzelnigg1985JCP,Bartlett1989CPL,Taube2006IJQC,Yanai2009IJQC,Yanai2009IJQC,Yanai2006JCP, Harsha2018JCP}---can be solved at a polynomial-scaling cost using a quantum computer~\cite{Peruzzo2014NatComm}.
A UCC wave function can be prepared on a quantum computer using the Trotter approximation with a polynomial number of quantum gates.
Although the gate count for the accurate UCC can be much larger than what today's quantum devices are capable of~\cite{Kuhn2019JCTC}, UCC can be a good starting point for analyzing the power of the quantum computer in the field of quantum chemistry.

The method enabling UCC on a quantum computer is called variational quantum eigensolver (VQE), which is a kind of quantum-classical hybrid algorithm~\cite{Peruzzo2014NatComm}.
In the VQE, a wave function is prepared through a parametrized quantum circuit corresponding to a wave function ansatz (e.g., UCC).
Then, we measure its energy for given circuit parameters.
The parameters of the circuit are iteratively tuned by a non-linear optimizer running on a classical computer to minimize the energy.
VQE calculations using actual quantum computers have already been performed for small molecules~\cite{Peruzzo2014NatComm, OMalley2016PRX, Kandala2017Nature, Hempel2018PRX, Ryabinkin2018JCTC2, McCaskey2019NPJQ, Gao2019arXiv}.
Recently, researchers have proposed electron-correlation methods based on UCC for quantum computers~\cite{Ryabinkin2018JCTC, Lee2018JCTC2, Grimsley2019NatComm, Huggins2019arXiv, Bauman2019JCP, Bauman2019JCP2, Ryabinkin2019iQCC, Matsuzawa2020JCTC, Greene2019arXiv, Evangelista2019JCP}.

Although VQE allows the determination of UCC parameters based on the Rayleigh--Ritz variational procedure, the obtained UCC wave functions are not fully variational.
UCC and its variants employ a Hartree--Fock determinant as a reference wave function; the orbitals are fixed and not altered during a UCC calculation.
However, it is well-known that the Hartree--Fock orbitals are not optimal orbitals for a correlated wave function.
One method for obtaining such optimal orbitals is to optimize orbitals in such a way that the gradients of the energy with respect to orbital rotation parameters vanish.
The combination of CC with orbital optimization (OO-CC) was first briefly mentioned in the paper of Purvis and Bartlett and then introduced by Sherrill et al.~\cite{Purvis1982JCP, Sherrill1998JCP}.
Since then, orbital optimized coupled cluster doubles (OO-CCD) and its variants have been developed by various researchers~\cite{Sherrill1998JCP,  Kohn2005JCP, Lochan2007JCP, Neese2009JCTC, Kollmar2010TCA, Bozkaya2011JCP, Bozkaya2012JCP, Robinson2012JCP, Robinson2013JCP, Lan2013JCP, Bozkaya2013JCP, Boguslawski2014JCP, Bozkaya2016PCCP, Kats2018JCTC, Lee2018JCTC, Myhre2018JCP, Tew2018JCTC, Sato2018JCP}.

This article concerns the orbital-optimization technique to UCC in the context of VQE.
At this moment, the size of the orbital space that can be handled by a quantum computer is severely limited because of the number of the available qubits.
Therefore, active space approximation is indispensable when we wish to use a quantum computer for quantum chemical problems.
Improvement of the active space can be achieved by optimizing molecular orbitals with the VQE, which leads to the reduced number of qubits.
Furthermore, the orbital-optimized VQE (OO-VQE) is a fully variational method, and the molecular gradients of OO-VQE (e.g., forces) can be calculated without solving response equations.
The idea of using the orbital-optimization techniques for quantum computers has been already reported by Reiher et al.\ for the phase estimation algorithm (PEA)~\cite{Reiher2017PNAS} and by Takeshita et al.\ for the VQE~\cite{Takeshita2020PRX}.

In this study, we implement OO-VQE using a quantum circuit simulator; we propose an orbital optimized unitary coupled cluster doubles (OO-UCCD) as a wave-function model for OO-VQE. 
Note that after posting the initial manuscript of this work on the arXiv preprint server, an implementation of OO-UCCD by Sokolov et al. was appeared there~\cite{Sokolov2019arXiv}. 
Their study carefully performed the cost and accuracy analysis of several variants of OO-UCCD methods, while our work has focused on it's fully variational nature and computed analytical derivatives for geometry optimizations which is a vital part of quantum chemical calculations. To the best of our knowledge, this is the fist time geometry optimizations have been performed for polyatomic molecules using analytic first derivatives of the VQE and {\it ab initio} Hamiltonian.

The remainder of this paper is organized as follows: First, Sec.\ II following chapter describes theory of orbital optimized UCC (OO-UCC) based on VQE.
Sec.\ III provides a brief description of the implementation of OO-UCC using a quantum circuit simulator.
Sec.\ III also discusses simple numerical experiments to exhibit its usefulness.
Finally, Sec.\ IV concludes the paper.

\section{\label{sec:theory} Theory}

\subsection*{Unitary coupled cluster}
The molecular electronic Hamiltonian in a spin-free form is expressed as 
\begin{align}
    \hat{H} = \sum_{p,q} h_{pq} \hat{E}_{pq} + \sum_{p,q,r,s} h_{pqrs} \{ \hat{E}_{pq} \hat{E}_{rs} 
    - \delta_{qr} \hat{E}_{ps} \} \label{eq:electron_hamiltonian},
\end{align}
where $h_{pq}$ and $h_{pqrs}$ are one- and two-electron integrals, respectively.
$\hat{E}_{pq}$ is a singlet excitation operator and is defined as $\hat{E}_{pq}=\hat{c}^\dagger_{p,\alpha} \hat{c}_{q,\alpha} + \hat{c}^\dagger_{p,\beta}　\hat{c}_{q,\beta}$, where $\hat{c}^\dagger_{p,\alpha}$ and $\hat{c}_{p,\beta}$ are creation and annihilation second quantized operators, respectively.
$p, q, r, s$ are the indices of general molecular spatial orbitals.

A wave function in the traditional coupled cluster ansatz is given as
\begin{align}
    \ket{\Psi} = e^{\hat{T}} \ket{0},
\end{align}
where $\hat{T}$ is an excitation operator $\hat{T}=\hat{T}_1+\hat{T}_2+\hat{T}_3+\cdots$ and $\ket{0}$ is a reference wave function.
In contrast, UCC uses an anti-hermite operator $\hat{A}$ defined by the difference of the amplitude operator $\hat{T}$ of TCC and its Hermitian conjugate, i.e.,  $\hat{A}=\hat{T}-\hat{T}^\dagger$.
Therefore, a wave function of the UCC ansatz is expressed as, 
\begin{align}
    \ket{\Psi} = e^{\hat{A}} \ket{0}.
\end{align}
The Baker--Campbell--Hausdorff (BCH) expansion of the similarity transformed Hamiltonian of the traditional CC is terminated at the finite order, whereas that of UCC is not owing to de-excitation operators $\hat{T}^\dagger$.
The infinite BCH expansion makes the implementation of UCC on a classical computer unfeasible. 

\subsection*{Orbital optimization}

Optimizing orbitals is equivalent to minimizing a wave function with respect to orbital rotation parameters $\kappa$.
The energy function of OO-UCC is given by
\begin{align}
    E(A, \kappa) = \braket{\Psi | e^{-\hat{\kappa}} \hat{H} e^{\hat{\kappa}} | \Psi} = \braket{0 | e^{-\hat{A}} e^{-\hat{\kappa}} \hat{H} e^{\hat{\kappa}} e^{\hat{A}} | 0},
    \label{eq:orbital-optimization}
\end{align}
where the orbital rotation operator is defined as $\hat{\kappa} = \sum_{pq} \kappa_{pq} (\hat{E}_{pq}-\hat{E}_{qp})$.
When UCC parameters $A$ are fixed, the second order expansion of the energy function becomes
\begin{align}
E(A, \kappa) \approx &
    \braket{\Psi | \hat{H} | \Psi} 
    + \sum_{pq} \kappa_{pq} \braket{\Psi | [\hat{H}, \hat{E}^{-}_{pq} ] | \Psi}  \nonumber \\
&+ \frac{1}{2} \sum_{pq,rs}
    \kappa_{pq}
    \bra{ \Psi } [[\hat{H}, \hat{E}^{-}_{pq}], \hat{E}^{-}_{rs}]  \nonumber \\
& +[[\hat{H}, \hat{E}^{-}_{rs}], \hat{E}^{-}_{pq} ] \ket{ \Psi}  \kappa_{rs},
\end{align}
where $\hat{E}^{-}_{pq} = \hat{E}_{pq}-\hat{E}_{qp}$.  
By taking the derivative with respect to $\kappa$, the following Newton--Raphson equation is obtained:
\begin{align}
    \mathbf{H}\mathbf{\kappa} = -\mathbf{g},
    \label{eq:oo-nr}
\end{align}
whose elements are  
\begin{align}
    H_{pq,rs}
    = \frac{1}{2}\braket{\Psi|[[\hat{H},\hat{E}^{-}_{pq}],E^{-}_{rs}] 
    +[[\hat{H},\hat{E}^{-}_{rs}],\hat{E}^{-}_{pq}]|\Psi} 
\end{align}
\begin{align}
    g_{pq} &= \braket{\Psi|[\hat{H},\hat{E}^{-}_{pq}]|\Psi}.
\end{align}
$H$ and $g$ are often called electronic Hessian and gradients, respectively.
One-particle and two-particle reduced density matrices (1RDM and 2RDM) are required to compute them in addition to the molecular Hamiltonian integrals $h_{pq}$ and $h_{pqrs}$.
They are readily available in VQE, because it measures 1RDM and 2RDM to compute electronic energy in a given quantum circuit. 

\subsection*{Orbital optimized unitary coupled cluster doubles}

The UCC singles and doubles (UCCSD) is given as 
\begin{align}
    \ket{\Psi^{\rm{UCCSD}}} = e^{\hat{A}_1+\hat{A}_2} \ket{0},
    \label{eq:UCCSD}
\end{align}
where $\hat{A}_n=\hat{T}_{n}-\hat{T}^{\dagger}_{n}$ consists of $n$-excitation operators $\hat{T}_{n}$ and their conjugates.
Starting from the UCCSD ansatz Eq.\ \ref{eq:UCCSD}, we consider the following wave-function model by separating the singles and doubles parts:
\begin{align}
    \ket{\Psi^{\rm{UCCSD'}}} 
    = e^{\hat{A}_2} e^{\hat{A}_1} \ket{0}.
    \label{eq:UCCSD_decomposed}
\end{align}
The UCC operator $\hat{A}$ is not commutable unlike TCC, because of the existence of de-excitation operators $\hat{T}^\dagger$.
Therefore, the decomposed UCCSD ansatz is different from the original ansatz.
The singles part $e^{\hat{A}_1}$ in this model is identical to the orbital rotation unitary operator $e^{\kappa}$ appeared in Eq.\  \ref{eq:orbital-optimization}.
This implies that we can optimize its singles part $e^{\hat{A}_1}$ variationally using a classical computer via the well-established orbital-optimization technique.
The singles only alter the Hartree--Fock determinant to another determinant $|\tilde{0}\rangle = e                                                                             ^{\hat{A} _1} |0\rangle$.

Considering the Slater determinant $|\tilde{0}\rangle$ as a reference wave function for UCC, we rewrite Eq.\ \ref{eq:UCCSD_decomposed} and propose the orbital-optimised unitary coupled cluster doubles (OO-UCCD) model, given as 
\begin{align}
    \ket{\Psi^{\rm{OO-UCCD}}}
    = e^{\hat{\tilde{A}}_2} \ket{\tilde{0}}.
    \label{eq:OO-UCCD}
\end{align}
The doubles part $e^{\hat{\tilde{A}}_2}$ in Eq.\ \ref{eq:OO-UCCD} is optimized by the VQE, while the reference determinant (i.e., the singles) is optimized by a classical computer using 1RDM and 2RDM from VQE.
In practice, we repeatedly perform the VQE and the orbital optimization until convergence.
In exchange for self-consistency, Eq.\ \ref{eq:OO-UCCD} requires less complicated quantum circuit than Eq.\ \ref{eq:UCCSD_decomposed}.

\subsection*{Analytical first derivatives of energy}

An important advantage of OO-VQE including OO-UCCD is that all the wave-function parameters are variationally determined. This feature allows us to compute first analytical derivatives of the energy without solving any additional equation. They are vital quantities for quantum chemical calculations, because static molecular properties such as forces on nuclei are defined as the derivatives of the energy with respect to external parameters $x$ (such as a position of atom or an external electric field). Here the energy is considered as a function of external parameters $x$, the VQE circuit parameters $\theta$ and the orbital parameters $kappa$, denoted as $E(x,\theta, \kappa)$ in this subsection.  Then, the first derivatives of the VQE energy are given as
\begin{align}
\frac{dE(x,\theta, \kappa)}{dx}
&=\frac{\partial E(x,\theta, \kappa)}{\partial x} \nonumber \\
&+\frac{\partial E(x,\theta, \kappa)}{\partial \theta}
\frac{\partial \theta}{\partial x}
+\frac{\partial E(x,\theta, \kappa)}{\partial \kappa}
\frac{\partial \kappa}{\partial x}, 
  \label{eq:first-deriv}
\end{align}
where $\theta$ and $\kappa$ indicate the quantum circuit parameters and molecular orbital parameters, respectively.
The first term of the right hand side of Eq.\ \ref{eq:first-deriv} is the Hellmann-Feynman term and equivalent to the expectation value of the derivative of the Hamiltonian.
The second term is zero since the VQE variationally determines $\theta$.
In the same way, the third term is also zero when the orbital is optimized.
Thus, the first-order OO-VQE energy derivatives can be determined by evaluating the expectation value of the derivative of the Hamiltonian.

This is not the case for the standard VQE without $\rm OO$, which needs the third term to compute the first derivatives.
To obtain the orbital response ${\partial \kappa}/{\partial x}$ in that term, we must solve the following first-order coupled-perturbed Hartree-Fock (CPHF) equation: 
\begin{equation}
    \frac{\partial^2 E_{\rm HF}(x,\kappa)}{\partial \kappa \partial \kappa} \frac{\partial \kappa}{\partial x}
    = \frac{\partial^2 E_{\rm HF}(x,\kappa)}{\partial x \partial  \kappa },
    \label{eq:CPHF}
\end{equation}
where $E_{\rm HF}(x,\kappa)$ is the Hartree--Fock energy.
In practice, we use so-called the Z-vector technique~\cite{Handy1984JCP,Helgaker1989TCA} so that we can avoid to solve Eq.\ \ref{eq:CPHF} for each of the external parameters (e.g. each Cartesian coordinate of a molecule)~\cite{Parrish2019arXiv}.
The second or higher-order derivatives require not only the orbital response but also the circuit parameter response.
The latter can be computed analytically on a quantum computer using a method recently developed by a part of authors of this study~\cite{Mitarai2020PRR}.  

\subsection*{Trotterization and Brueckner orbitals}

Translating the UCC generator $e^{\hat{A}}$ into a quantum circuit 
needs Trotterization, which introduces the error owing to the finite Trotter number.
Hereafter, we denote UCCSD and UCC doubles (UCCD) approximated by the $n$-step Trotter expansion as ${\rm UCCSD}_{n}$ and ${\rm UCCD}_{n}$, respectively.
The wave-function ansatz, which we implement in this study, corresponds to ${\rm OO\text -UCCD}_{1}$, where the Trotter expansion is truncated at the very first step. 
The ${\rm OO\text -UCCD}_{1}$ ansatz can be written as
\begin{align}
    \ket{\Psi}
    &= \prod_{\mu} \left( e^{\hat{\tilde{A}}_{2,\mu}} \right) e^{\hat{A}_{1}}\ket{0} \nonumber \\
    &= \prod_{\mu} \left( e^{\hat{\tilde{A}}_{2,\mu}} \right) \ket{\tilde{0}},
    \label{eq:uccdn1}
\end{align}
where $\mu$ is an index for each double excitation (and de-excitaion) operator and 
$\hat{A}_{2}=\sum_{\mu}\hat{A}_{2,\mu}$.
Although a single-Trotter-step UCC ansatz appears to be a crude approximation, Barkoutsos et al.\ have shown that it actually reproduces ground-state energy accurately~\cite{Barkoutsos2018PRA}.
O'Malley et al.\ have pointed out that the variational flexibility allows such an approximated wave function model to absorb the Trotterization error~\cite{OMalley2016PRX}.

A notable feature of the ${\rm OO\text -UCCD}$ ansatz is that the variational orbitals coincide with the so-called Brueckner orbitals. 
Brueckner orbitals are optimal orbitals for a correlated wave function, where the singles' contribution (i.e., $\hat{T}_{1}$ or $\hat{A}_{1}$) vanishes.
It is known that, in the TCC framework, the variationally optimized orbitals are not the same as Brueckner orbitals.
This is because of the difference between $\hat{T}_{1}$ and the orbital rotation operators.
On the other hand, the singles of UCC $\hat{A}_{1}$ are identical to them.
Nonetheless, the Brueckner orbitals of UCCSD are not identical to the OO-UCCSD optimzied orbitals because the anti-hermitian operators $\hat{A}_{1}$ and $\hat{A}_{2}$ are not generally commutable to each other and the Trotterization makes a difference.
The OO-UCCD ansatz is based on the separation of the singles and doubles {\it a posteriori} as shown in Eq.\ \ref{eq:UCCSD_decomposed}.
Therefore, the OO-UCCD naturally satisfies the Brueckner condition $e^{\hat{\tilde{A}}_1}=0$ and the variational condition $\frac{\partial E}{\partial \kappa}=0$, simultaneously: variational orbitals are Brueckner orbitals in this ansatz.

\section{\label{sec:results} Numerical examples}
In this section, we present some numerical results using the proposed method.
Computational details are as follows.
We have implemented OO-UCCD in Python using Qulacs{Qulacs}, PySCF~\cite{PYSCF}, and OpenFermion~\cite{Openfermion} program packages.
Qulacs is used to simulate quantum circuits.
PySCF is used for orbital optimizations and for evaluating molecular Hamiltonian integrals~\cite{Sun2017CPL}.
OpenFermion is employed for mapping molecular Hamiltonian into a quantum circuit based on the Jordan--Wigner transformation.

\subsection*{Without the active space approximation}

In this subsection, we show the results without the active space approximation.
It means that all the orbitals were mapped into qubits.
Then, the results with active space approximation will be illustrated in the next subsection.
Hereafter, we denote the $\rm OO\text -UCCD$ and $\rm UCCSD$ using the active space approach as $\rm AS\text -\rm OO\text -UCCD$ and $\rm AS\text -\rm UCCSD$, respectively, to distinguish them from no-active-sapce calculations.
On top of that, the active space sizes are explicitly written after the name of an electron correlation methods whenever the approximation is used.
Say, $\rm OO\text -UCCD$ and MP2 with the active space consisting of 6 spatial orbitals and 4 electrons are written as ${\rm AS\text -\rm OO\text -UCCD}(6o,4e)$ and MP2$(6o,4e)$, respectively.  

\begin{figure}
    \includegraphics[width=0.92\linewidth]{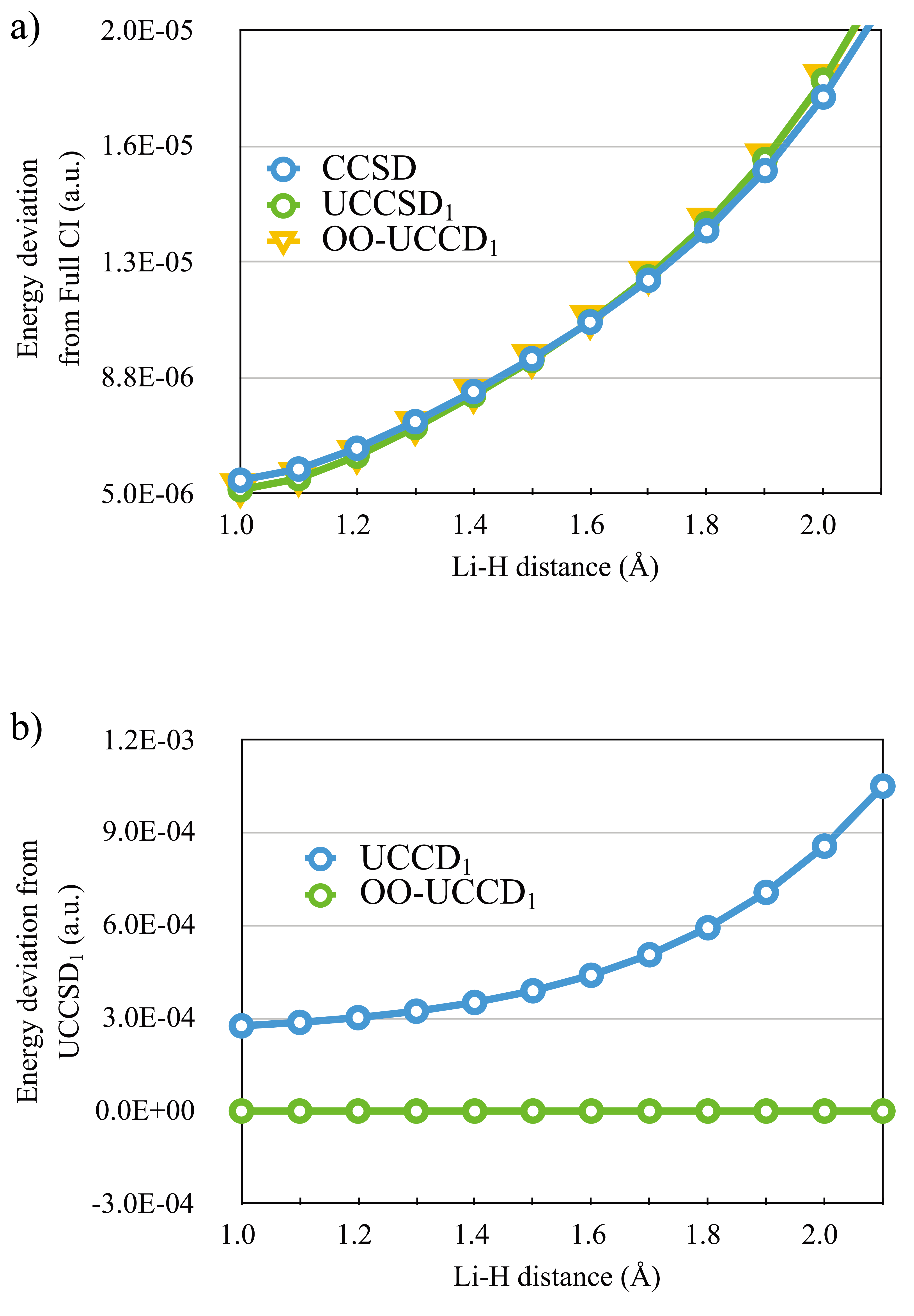}
    \caption{
     Error of the ${\rm OO\text -UCCD_{1}}$ method along with potential energy curves of the LiH molecule using the STO-3G basis sets.
     a) Deviations from FCI energies compared with ${\rm UCCSD}_{1}$ and the standard CCSD method.
     b) Deviations from ${\rm UCCSD_{1}}$ energies compared with ${\rm UCCD}_{1}$. 
    }
    \label{fig:LiH-STO3G}
\end{figure}

 We first investigate the potential energy curve (PEC) of LiH.
 Figure.\ \ref{fig:LiH-STO3G} shows the errors of our method with respect to the PECs computed by FCI and ${\rm UCCSD}_{1}$ with the STO-3G basis sets.
 We did not employ the active space approximation.
 It can be seen that the energy difference between ${\rm UCCSD}_{1}$ and ${\rm OO\text -UCCD}_{1}$ is notably small in  the entire range of the PEC.
 The energy deviation of ${\rm OO\text -UCCD}_{1}$ from ${\rm UCCSD}_{1}$ is $10^{-7}$mhartree at 1.3 \AA\ Li-H distance, whereas at 2.1 \AA\ it is $3 \times 10^{-5}$ mhatree.
 In contrast, the deviation of ${\rm UCCD}_{1}$ is at least five orders of magnitude larger than ${\rm OO\text -UCCD}_{1}$.
 ${\rm OO\text -UCCD}$ is as accurate as ${\rm UCCSD}$; likewise the standard OO-CCD calculations reproduce the CCSD results well.
 This indicates that ${\rm UCCSD}$-level results can be obtained with a shallower quantum circuit using the orbital optimization technique at the cost of repeated VQE optimizations. 
 Note that Trotterization introduces dependency on the operator ordering.
 Although Grimsley et al.\ have illustrated that the effect of the operator ordering could be significant, Grimsley's results suggest that the impact of the ordering is not important on the systems used in this study~\cite{Grimsley2019JCTC}.


\begin{table}
\caption{ RMSD errors (in angstrom) for geometries optimized by OO-UCCD$_1$/STO-3G and standard wave-function models HF, MP2, CCSD and CCSD(T) relative to FCI. The parentheses show deviations of energies at optimized geometries from FCI.}
\label{tab:geom-opt}
\begin{tabular}{ll cc cc cc}
\hline \hline
 & & NH$_3$ &  &  & H$_2$O &  \\
\hline
HF            & &3.2E-02 (7.3E-02) & & &2.4E-02  (5.7E-02) \\
MP2           & &9.0E-03 (2.1E-02) & & &9.3E-03  (1.7E-02) \\
CCSD          & &2.9E-04 (2.5E-04) & & &1.7E-04  (1.5E-04) \\
UCCD$_1$      & &5.1E-04 (3.7E-04) & & &5.1E-04	 (4.1E-04) \\
UCCSD$_1$     & &3.3E-04 (1.8E-04) & & &7.1E-05  (1.0E-04) \\
OO-UCCD$_1$   & &3.3E-04 (1.8E-04) & & &6.7E-05  (1.0E-04) \\
CCSD(T)       & &8.1E-05 (1.3E-04) & & &1.7E-04  (7.8E-05) \\
\hline \hline
\end{tabular}
\end{table}

Next, we report the geometry optimization of the water and ammonia molecules with STO-3G basis sets using the analytical derivatives of ${\rm UCCD}_{1}$, ${\rm UCCSD}_{1}$ and ${\rm OO\text -UCCD}_{1}$.
For comparison, we carried out Hartree--Fock, MP2, CCSD, CCSD(T), and FCI calculations.
The CCSD and CCSD(T) geometries were obtained by the ORCA program package using numerical gradients~\cite{Neese2012WIRCMS,Neese2018WIRCMS}.
All electrons were correlated in these calculations.

Table\ \ref{tab:geom-opt} shows the error of the total energy from FCI for each optimised geometry, and the root mean square deviation (RMSD) error of each optimized structure. 
The RMSD error is a commonly-used measure of the structural difference of two molecules. It was computed using the deviations of Cartesian coordinates of atoms from the reference FCI values, where the molecular structure was superimposed on the reference.
It shows that ${\rm OO\text -UCCD}_{1}$ is close to FCI in terms of both structure (i.e., $3\times 10^{-4}$ \AA ) and energy (i.e., 0.2 mhartree). 
One can seen ${\rm OO\text -UCCD}_{1}$ and ${\rm UCCSD}_{1}$ provided virtually the same geometries and energies for these two systems.
On the other hand, UCCD$_1$ is less accurate than CCSD and the other two UCC models because of the lack of the singles contributions, 
though UCCD$_1$ is more precise than MP2.
The number of the VQE parameters $\theta$ and the depth of the quantum-circuits of ${\rm OO\text -UCCD}_{1}$ and UCCD$_1$ were 120 and 2720, respectively, for $\rm NH_3$; those of ${\rm UCCSD}_{1}$ were 135 and 2780.
These results exhibit that the orbital-optimization allows us to slightly reduces the circuit depth while retaining the accuracy.

Furthermore, the comparison with CCSD and CCSD(T) suggests that ${\rm OO\text -UCCD}_{1}$'s energy is more accurate than that of CCSD and less than that of CCSD(T) when electron correlation is weak.
This tendency is consistent with the findings of K\"uhn et al.\ for the total energy and reaction energy~\cite{Kuhn2019JCTC}; in terms of geometry, ${\rm OO\text -UCCD}_{1}$ is slightly better than both CCSD and CCSD(T) for $\rm H_2O$, while for $\rm NH_3$ it is close to CCSD and worse than CCSD(T).

\subsection*{With the active space approximation}

\begin{figure}
    \includegraphics[width=0.85\linewidth]{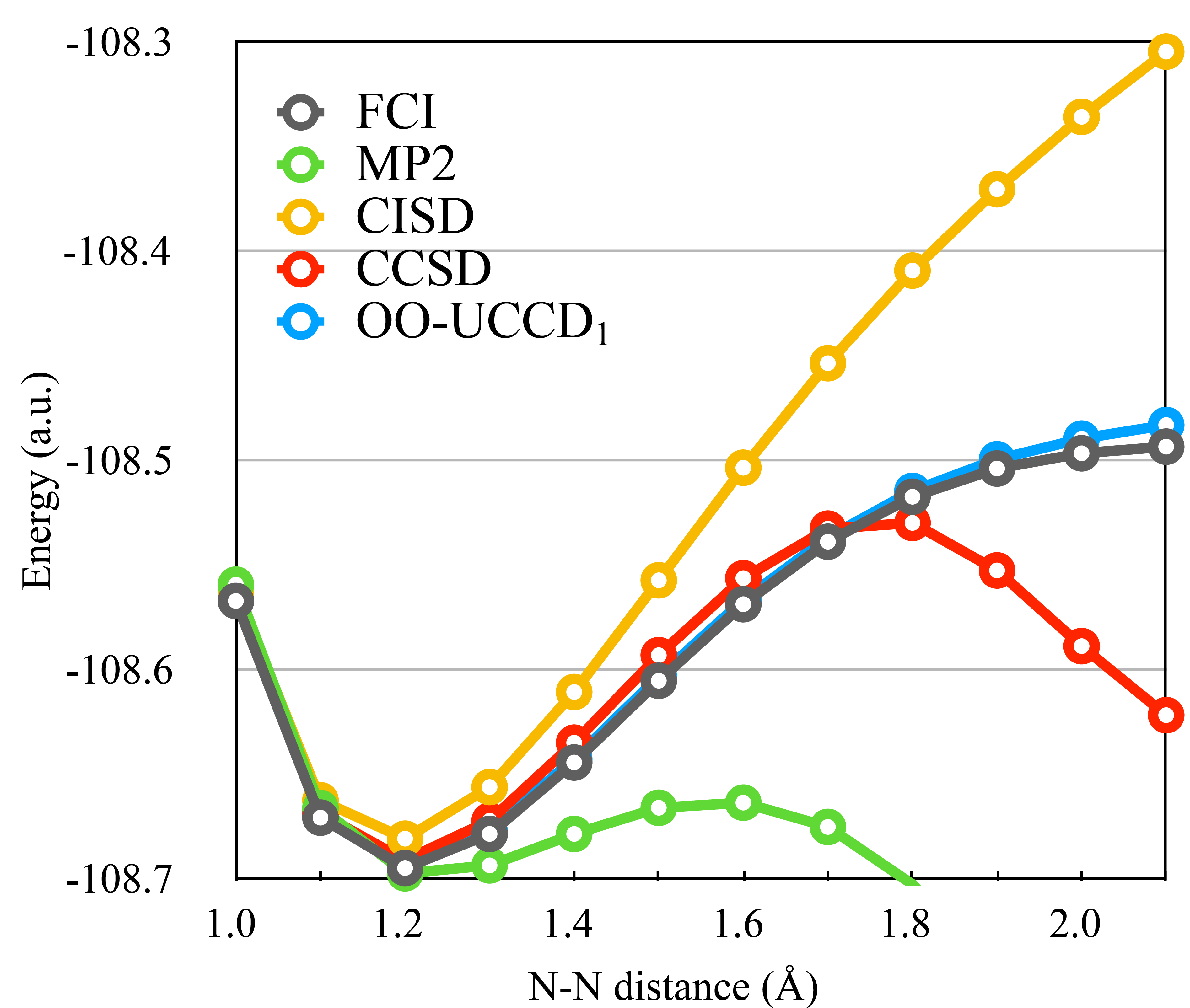}
    \caption{
     Potential energy curves of the $\rm{N}_{2}$ molecule computed at  ${\rm AS\text -\rm OO\text -UCCD_{1}}(6o,6e)$, MP2$(6o,6e)$, CISD$(6o,6e)$, CCSD$(6o,6e)$ and FCI$(6o,6e)$ using the STO-3G basis sets, where six orbitals and six electrons were correlated.
     Other orbitals were kept fixed and not used during all the post-Hartree--Fock calculations; here, the external orbital rotation was omitted for ${\rm AS\text -\rm OO\text -UCCD_{1}}(6o,6e)$. 
    }
    \label{fig:N2-PEC}
\end{figure}

Next, we examine the potential energy curve (PEC) of N$_2$ using the active space approximation.
This system, involving triple bond breaking, is a well-known benchmark for electron correlation methods, where the standard methods of many body perturbation theory such as MP2, CCSD, and CCSD(T) breakdown.
We have computed the PEC at the ${\rm AS\text -\rm OO\text -UCCD_{1}}(6o,6e)$/STO-3G level of theory by fixing the lowest four occupied orbitals and eight electrons.
FIG.\ \ref{fig:N2-PEC} shows the PEC along with those computed by MP2$(6o,6e)$, CISD$(6o,6e)$, CCSD$(6o,6e)$, and FCI$(6o,6e)$.
It shows that AS-OO-UCCD can treat a multiple-bond-breaking system appropriately where electrons of the breaking chemical bond are strongly correlated. 
In this system, the differences among AS-UCCD, AS-OO-UCCD, and AS-UCCSD were small.
They are at most 0.2 kcal/mol owing to the little orbital relaxation effect in the small basis sets.
The root mean square deviations (RMSD) of ${\rm AS\text -\rm UCCD_{1}}(6o,6e)$ and ${\rm AS\text -\rm OO\text -UCCD}_{1}(6o,6e)$ with respect to ${\rm AS\text -\rm UCCSD}_{1}(6o,6e)$ PEC is 0.02 kcal/mol in the range of 1.0--2.1 \AA.

\begin{figure}
    \includegraphics[width=0.7\linewidth]{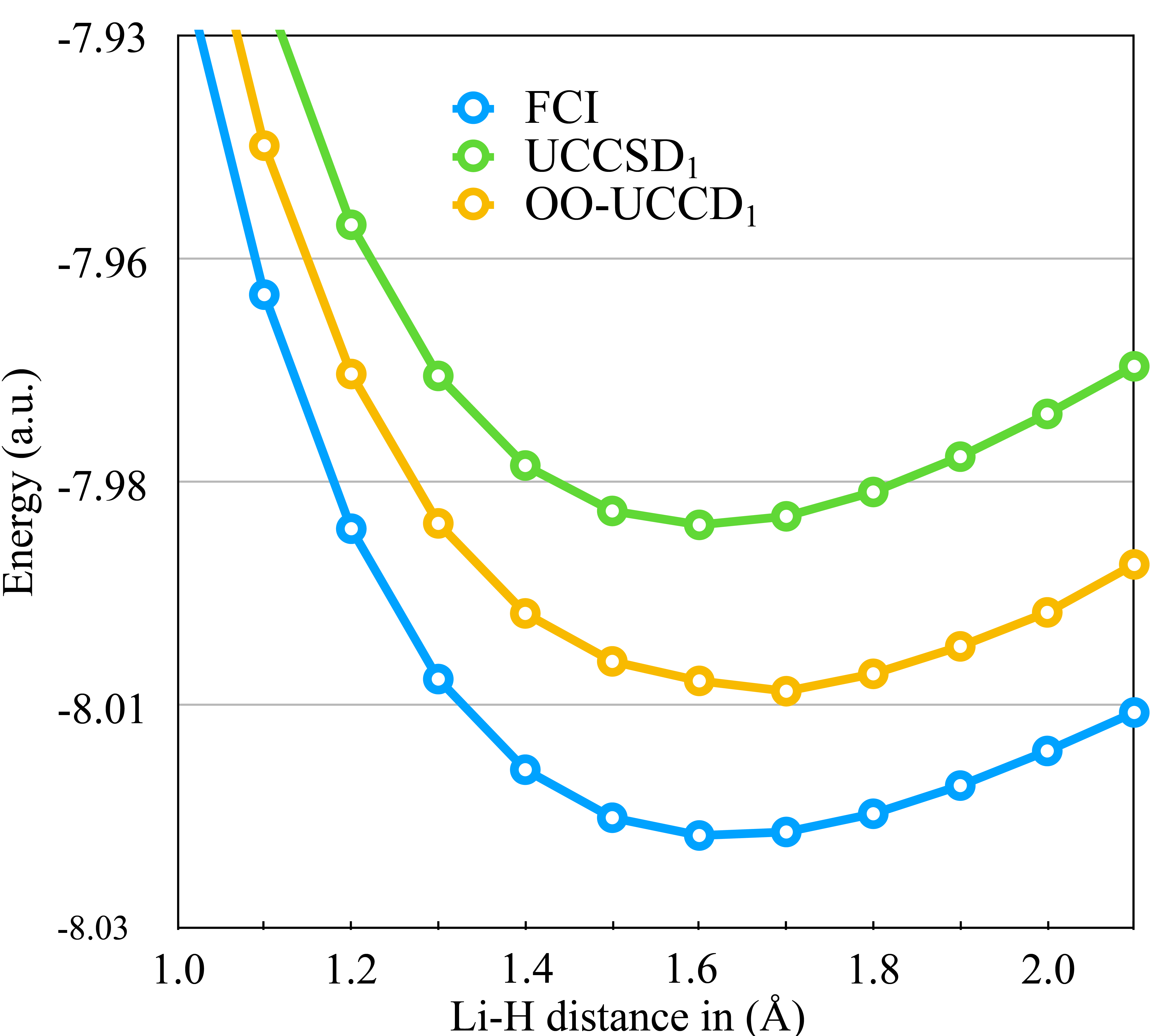}
    \caption{
     Potential energy curves of the LiH molecule computed at ${\rm AS\text - OO\text -UCCD_{1}}(4o,2e)$, ${\rm AS\text - UCCSD_{1}}(4o,2e)$, and FCI using the 6-311G basis sets.
     ${\rm AS\text -\rm OO\text -UCCD_{1}}$ and ${\rm AS\text -\rm UCCSD_{1}}$ calculations employed active space consisting of four orbitals and two electrons (i.e., $(4o,2e)$), while all the orbitals and electrons were correlated for FCI.   
    }
    \label{fig:LiH-6-311G}
\end{figure}

The PEC of LiH with 6-311G basis sets, given in FIG.\ \ref{fig:LiH-6-311G}, is another example.
For comparison, we computed the PEC by ${\rm AS\text -\rm OO\text -UCCD}_{1}(4o,2e)$ and also by ${\rm AS\text -\rm UCCSD}_{1}(4o,2e)$ and FCI.
Only four orbitals and two electrons were correlated in those VQE calculations, and the orbital optimization was performed for all the orbitals and electrons.
This example shows how the orbital optimization effectively considers the electron correlations outside the active space.
Figure\ \ref{fig:LiH-6-311G} illustrates that ${\rm AS\text -\rm OO\text -UCCD_{1}}(4o,2e)$ has lower enery and is closer to FCI without active space than ${\rm AS\text -\rm UCCSD}(4o,2e)$, indicating that ${\rm AS\text -\rm OO\text -UCCD_{1}}$ captures more electron correlation effects. 


\section{\label{sec:summary} Summary}
In this work, we have developed OO-UCCD.
OO-UCCD treats singles contributions not on quantum computers but on classical computers.
This slightly reduces the number of gates and the depth in the UCCSD quantum circuit as noted by Takeshita and his coworkers~\cite{Takeshita2020PRX},
while the conventional OO-CC simplifies energy expressions and amplitude equations.
Moreover, all the wave function parameters are fully variationally determined in OO-UCC.
This property makes the time-independent first-order properties readily available.
Therefore, geometry optimisation or {\it ab initio} molecular dynamics can be performed using VQE without solving orbital response equations.
These aspects seem useful, especially in the age of noisy intermediate-scale quantum computers (NISQ)~\cite{Preskill2018Quantum,Arute2019Nature}, where the number of qubits and the coherence time are severely limited.
Such an OO-VQE method may be useful for solving quantum chemical problems once quantum computers become commonplace.

\section*{\label{sec:acknowledements} Acknowledgements}
This work was supported by MEXT Quantum Leap Flagship Program (MEXT Q-LEAP) Grant Number JPMXS0118067394, JPMXS0118068682.
WM wishes to thank JSPS KAKENHI No.\ 18K14181 and JST PRESTO No.\ JPMJPR191A.
KM was supported by JSPS KAKENHI No.\ 19J10978.
KM thanks for METI and IPA for their support through MITOU Target program.
Some calculations were performed using the computational facilities in the Institute of Solid State Physics at the University of Tokyo, and in Research Institute for Information Technology (RIIT) at Kyushu University, Japan.

\bibliography{oouccd}

\end{document}